\documentclass[epjCONF,columns]{svjour} %for 2 columns format
\usepackage{graphics}
\usepackage[varg]{txfonts} % Times fonts
\usepackage[latin1]{inputenc}

\session-title{Hadron Collider Physics Symposium 2011}
\begin{document}
\title{Search for supersymmetry in hadronic final states with $M_{T2}$}
\author{Hannsj\"org Weber\inst{1}\fnmsep\thanks{\email{hannsjorg.artur.weber@cern.ch}} \textit{on behalf of the CMS collaboration}  }% \and Second author\inst{2} \and ... }
%\author{Hannsj\"org Weber\inst{1} \textit{on behalf of the CMS collaboration}  }% \and Second author\inst{2} \and ... }

%
\institute{\inst{1} Institute for Particle Physics, ETH Z\"urich, 8093 Zurich, Switzerland }% \and the second here \and ...}
\abstract{
We present the results of a search for physics beyond the Standard Model (BSM) using data of 1.1 $\mathrm{fb^{-1}}$ integrated luminosity collected by the CMS experiment at the LHC. Fully hadronic final states were selected based on the ``stransverse" mass variable $M_{T2}$ and interpreted in various models of supersymmetry (SUSY). Two complementary analyses were performed targeting different areas of the SUSY phase space. All backgrounds were estimated using both simulation and data-driven methods. As no excess of events over the expected background was observed exclusion limits were derived.
} %end of abstract
\maketitle
\section{Introduction}
\label{sec.introduction}
We describe a search \cite{PAS-SUS-11-005} for physics beyond the Standard Model in pp collisions collected by the Compact Muon Solenoid (CMS) detector \cite{CMSpaper} at the Large Hadron Collider (LHC) at a centre-of-mass energy of 7 TeV. 
The results are based on a data sample of 1.1 $\mathrm{fb^{-1}}$ of integrated luminosity collected in 2011. 
We use the ``stransverse mass" variable $M_{T2}$ \cite{Ref.Lester:1999tx} to select new physics candidates out of fully hadronic events. We divide our search into two channels: one targets high squark and gluino masses with a high $M_{T2}$ cut, the other heavy squarks and light gluinos with a medium $M_{T2}$ cut but including a $b$-tag and high jet multiplicities.
\\
In the following we describe the properties of $M_{T2}$ in sec. \ref{sec.MT2variable}, our analyis strategy and event selection in sec. \ref{sec.strategy.and.selection}, and the background estimation in sec. \ref{sec.bg.estimation}. 
In sec. \ref{sec.Results} we state the results of our search, and draw a conclusion in sec. \ref{sec.conclusion}.
\section{The search variable $M_{T2}$  }
\label{sec.MT2variable}
The variable $M_{T2}$ was introduced to measure the mass of primary pair-produced particles where both particles decay into detected and undetected particles (e.g. the lightest supersymmetric particle (LSP)). 
It is a generalization of the transverse mass $m_T$ in case of two identical decay chains each containing unobserved particles. 
The variable $M_{T2}$ is defined as
\begin{eqnarray}
 M_{T2}(m_{\chi}) = \min_{p_T^{\chi(1)} + p_T^{\chi(2)} = p_T^{miss}} 
  \left[ \max \left( m_T^{(1)} , m_T^{(2)} \right) \right] , %\nonumber
\label{eq.MT2.definition}
\end{eqnarray}
where $m_T$ is the transverse mass of the visible system and the corresponding LSP $\mathrm{\chi}$ of the decaying sparticle:
\begin{eqnarray}
  m_T^{(i)} = \sqrt{(m^{vis(i)})^2 + m_{\chi}^2
 + 2 \left( E_T^{vis(i)} E_T^{\chi(i)} - \vec{p}_T^{\; vis(i)} \cdot \vec{p}_T^{\; \chi(i)}
     \right)}. \nonumber
\label{eq.MT2.transmass}
\end{eqnarray}
In this analysis the stransverse mass $M_{T2}$ is not used for mass measurements but rather as a discovery variable \cite{Ref.Barr:2009wu}.

In order to associate all visible decay products to the decay chains of the two sparticles we cluster the jets of an event into two ``pseudojets" using a hemisphere algorithm described in \cite{ref.CMSTDR07}, Sect. 13.4. 
As seeds (inital axes) the direction of the two (massless) jets are chosen which have the largest invariant mass. 
We then associate a jet $k$ to the pseudojet $i$ rather $j$ if the Lund distance is minimal:
\begin{eqnarray}
 (E_i-p_i cos \theta_{ik}) \frac{E_i}{(E_i+E_k)^2} \leq  (E_j-p_j cos \theta_{jk}) \frac{E_j}{(E_j+E_k)^2} . \nonumber
%  \frac{(E_i-p_i cos \theta_{ik})E_i}{(E_i+E_k)^2} \leq   \frac{(E_j-p_j cos \theta_{jk})E_j}{(E_j+E_k)^2} . \nonumber
\label{eq.lunddistance}
\end{eqnarray}
\subsection{Advantages of $M_{T2}$  }
\label{sec.MT2advantages}
In order to gain a better understanding of the behaviour of $M_{T2}$ we take the case where we set all masses to zero and assume no initial-state radiation (ISR) or upstream transverse momentum\footnotemark[1]. 
\footnotetext[1]{Upstream transverse momentum is the transverse momentum which is not clustered to the pseudojets (e.g. jets outside of acceptance).}
In this simple case $M_{T2}$ becomes
\begin{eqnarray}
        \label{eq:MT2simplified}
 (M_{T2})^2  = 2  p_T^{vis(1)} p_T^{vis(2)} ( 1 + cos \phi_{12} ) ,
\end{eqnarray}
where $p_T^{vis(i)}$ is the transverse momentum of pseudojet $i$, and $\phi_{12}$ the angle between the two pseudojets in the transverse plane. 
It can be observed that for symmetric events ($p_T^{vis(1)} = p_T^{vis(2)}$) with large acoplanarity $M_{T2}$ behaves like the missing transverse momentum (MET). 
Thus SUSY with expected large MET will accumulate in the high $M_{T2}$ region. 
However, back-to-back systems or balanced events will populate the region with small $M_{T2}$. 
Thus $M_{T2}$ is robust against QCD jet mismeasurements: 
Mismeasurements along one of the pseudojets results in $M_{T2} \approx 0\mathrm{~GeV}$, while for asymmetric mismeasurements still $M_{T2} < \mathrm{MET}$.
\section{Analysis Strategy and Event Selection}
\label{sec.strategy.and.selection}
In this analysis we have established two search channels in order to be sensitive to different regions in the SUSY phase space. 
One approach, the High $M_{T2}$ analysis, targets events resulting from heavy sparticle production which is characterized by large MET and $M_{T2}$. 
The second approach, the Low $M_{T2}$ analysis, is designed to be sensitive to the region where squarks are heavy and gluinos relatively light. 
Here gluino-gluino production is dominant, the gluinos giving rise to three-body decays with small MET. 
Also, as stops and sbottoms are expected to be relatively light, these events can be enriched with $b$-quarks.
Thus the two strategies require two different sets of selection cuts stated in table \ref{tab.selection.highlowMT2}.
\begin{table}[h!]
\begin{center}
\caption{Event selection cuts which are specific to their strategies.}
\label{tab.selection.highlowMT2}       % Give a unique label
\begin{tabular}{|c|c|}
%\hline\noalign{\smallskip}
\hline
High $M_{T2}$ & Low $M_{T2}$ \\
\hline
at least 3 jets\footnotemark[2] & at least 4 jets \\
                      & at least 1 $b$-tag \\
$H_T > 600 \mathrm{~GeV}$\footnotemark[3] & $H_T > 650 \mathrm{~GeV}$ \\
$M_{T2} > 400 \mathrm{~GeV} $& $M_{T2} > 150 \mathrm{~GeV}$ \\
\hline
\end{tabular}
\end{center}
\end{table}
\footnotetext[2]{The jet selection requires $p_T > 20\mathrm{~GeV}, ~|\eta|<2.4$}
\footnotetext[3]{$H_T$ is the scalar sum of all jet-$p_T$.}
Besides this channel specific selection we also require:
\begin{itemize}
\item Lepton ($e,~\mu$) veto to reduce $W$+Jets and $t\bar{t}$ background. 
\item $\min\mathrm{\Delta\phi}(\mathrm{MET,~any~jet}) > 0.3$ to reduce further QCD.
\item $|\vec{MHT} - \vec{MET} | < 70 \mathrm{~GeV}$ to minimize the influence of unclustered energy (e.g. ISR) to the $M_{T2}$ shape\footnotemark[4].
\item MET tail cleaning cuts (e.g. noise filters) to filter out events with unphysical MET.
\end{itemize}
\footnotetext[4]{$\vec{MHT}$ is the negative vectorial sum of all jet-$p_T$.}
For the selection of data we require the data to pass $H_T$ trigger paths.
\section{Background Estimation Strategy}
\label{sec.bg.estimation}
For each type of background data-driven estimation methods have been designed: 
QCD is estimated from the bulk of the $M_{T2}$ distribution as described in sec. \ref{sec.QCDestimate}. 
In order to reduce the effect of signal contamination and statistical fluctuations the electroweak and top background is estimated from an adjacent control region in $M_{T2}$. 
The prediction is taken from data in the control region scaled by Monte-Carlo (MC) ratio of the event yield in the signal region over the yield in the control region. 
Similarly the uncertainties are scaled by a MC ratio. 
The control region for High $M_{T2}$ is defined as $200~\mathrm{GeV}<M_{T2}<400~\mathrm{GeV}$, for Low $M_{T2}$ it is defined as $100~\mathrm{GeV}<M_{T2}<150~\mathrm{GeV}$.
\subsection{QCD background estimation}
\label{sec.QCDestimate}
The QCD estimation method is based on the two variables $M_{T2}$ and $\mathrm{\Delta\phi}_{\min} = \min\mathrm{\Delta\phi}(\mathrm{MET,~any ~jet})$. These two variables are strongly correlated, but a factorization method can be applied if the functional form for the ratio $r(M_{T2}) = N(\mathrm{\Delta\phi}_{\min} \geq 0.3) / N(\mathrm{\Delta\phi}_{\min} \leq 0.2)$ is known. From simulation studies it has been found that for $M_{T2} > 50~\mathrm{GeV}$ the ratio falls exponantially: %(see fig. \ref{fig.QCDratio}):
\begin{eqnarray}
 r(M_{T2}) = \frac{N(\mathrm{\Delta\phi}_{\min} \geq 0.3)}{N(\mathrm{\Delta\phi}_{\min} \leq 0.2)}
  = \exp(a - b \cdot M_{T2}) + c
\label{eq.bkgdpred.ratio}
\end{eqnarray}
This behaviour was confirmed in data. 
The estimate has been performed by a fit from data in the QCD dominated region of $50~\mathrm{GeV}<M_{T2}<80~\mathrm{GeV}$ to extract the parameters $a$ and $b$. 
In order to get also parameter $c$ from data its value was fixed to the value of the ratio at $M_{T2} = 200~\mathrm{GeV}$ where the ratio still falls exponentially.
\subsection{$Z \to {\nu\nu}$ background estimation}
\label{seq.Znunu.estimate}
$Z \to {\nu\nu}$ is an irreducible background because the produced neutrinos leave the detector unmeasured and thus generate real MET. 
%This background has been estimated from the number of $W$+Jets in the muon channel. 
The number of $Z \to {\nu\nu}$+Jets events passing the event selection can be estimated from $W \to {\mu\nu}$+Jets via
\begin{eqnarray}
N_{Z\nu\nu}(est) = W(\mu\nu) \cdot  \frac{1}{\epsilon_{acc}\epsilon_{reco/iso}} \cdot  R_{ZW}
\label{eq.bkgdpred.ZnnWln}
\end{eqnarray}
where $W(\mu\nu)$ is the number of  $W \to {\mu\nu}$ events passing the event selection with additionally requiring one muon, $R_{ZW}$ is the ratio of $Z \to {\nu\nu}$ events to $W \to {\mu\nu}$ events, $\epsilon_{acc}$ is the acceptance, and $\epsilon_{reco/iso}$ is the combined reconstruction and isolation efficiency.
In order to reduce the $t\bar{t}$ background in the $W \to {\mu\nu}$ selection a $b$-tag veto has been applied, while the residual $t\bar{t}$ background has been estimated from the $b$-tagged region. Furthermore $\epsilon_{reco/iso}$ has been calculated from Tag \& Probe studies on $ Z \to {ll}$ events while the acceptance and the ratio $R_{ZW}$ are taken from MC.
\subsection{$W$ and Top background estimation}
\label{sec.W.Top.estimate}
The background due to $W$ and Top has two sources: %\\
Either a lepton ($e$, $\mu$) from a $W$ has been unobserved due to acceptance cuts, or has been ``lost" due to failing either identification, isolation, or reconstruction criteria. %\\
The other source are $W$ decays into neutrinos and taus which decay hadronically.
\\

The number of events with a  ``lost" lepton has been estimated from the number of events with one lepton found in data. 
This number is then corrected for the probability to loose a lepton via the formula 
\begin{eqnarray}
N^{pass \ veto}_{e, \mu} &=& (N_{e, \mu}^{reco} - N^{bg}_{e, \mu})\frac{1- \varepsilon_{e, \mu}}{\varepsilon_{e, \mu}},
\label{eq.bkgpred.LostLepton}
\end{eqnarray}
where $N_{e, \mu}^{reco}$ is the number of events containing a lepton, $N^{bg}_{e, \mu}$ is the expected background from processes other than $W$ or Top, and $\varepsilon_{e, \mu}$ is the probability for a $W\to l \nu$ ($l$ = $e ,\mu$, or $\tau\to e,\mu$)  passing all selection and reconstruction cuts.

The number of events with hadronic tau decays are taken from simulation and validated by data: 
The  $W\to l \nu$ ($l$ = $e ,\mu, \tau$) kinematics in simulation has been validated in data with one muon. 
Furthermore, it has been shown that tau decays are well modelled in the simulation \cite{PFT-11-001} justifying the use of simulation.
\section{Results}
\label{sec.Results}

\begin{figure}[h!]
\begin{center}
\resizebox{0.67\columnwidth}{!}{%
 \includegraphics{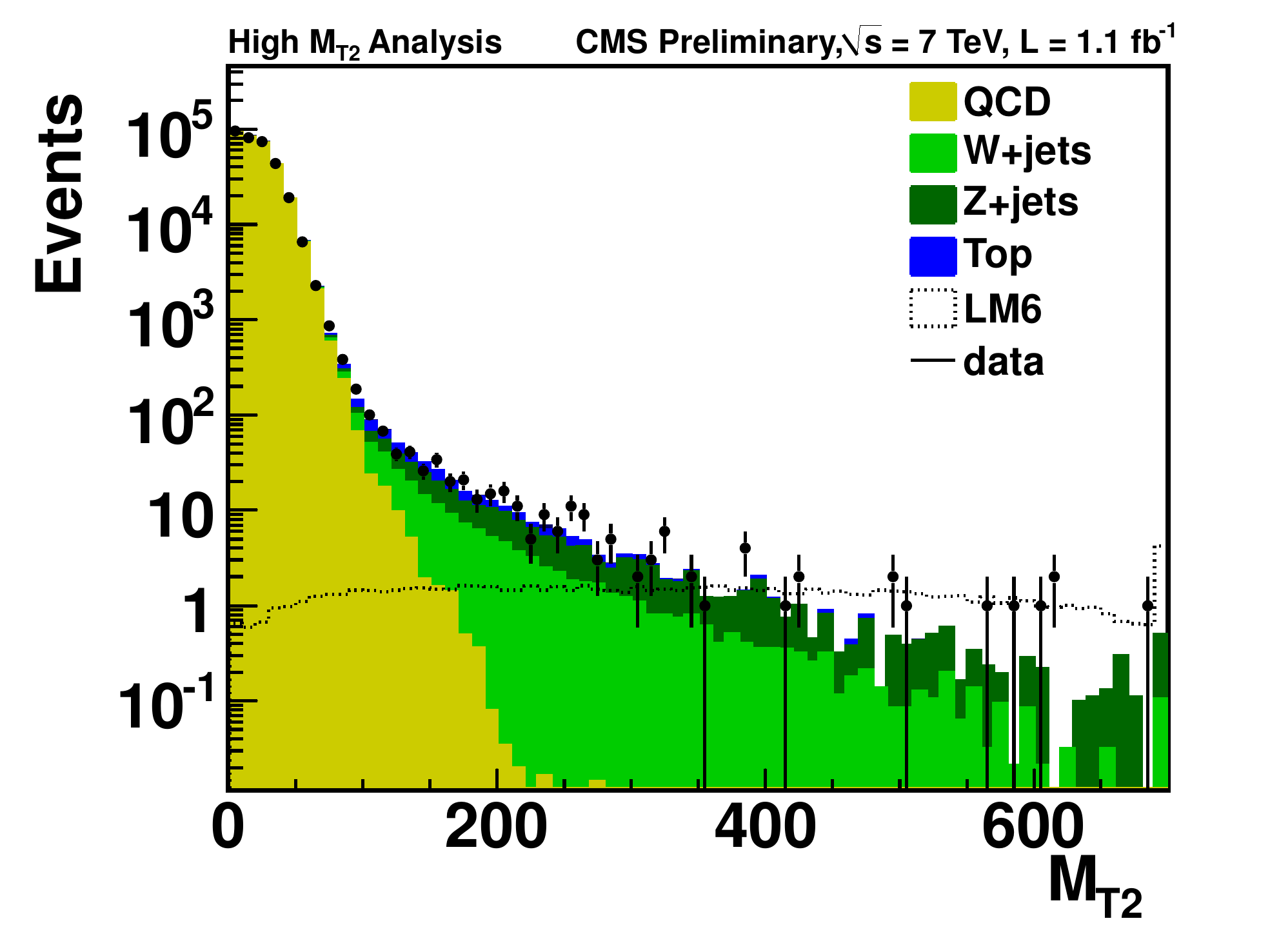} }
\caption{$M_{T2}$ distribution for the High $M_{T2}$ analysis. The MC background is normalized to 1.1 $\mathrm{fb}^{-1}$. A possible SUSY signal (LM6) is overlayed. Data are shown as dots on top of the background.  }
\label{fig.highMT2}       % Give a unique label
\end{center}
\end{figure}

\begin{figure}[h!]
\begin{center}
\resizebox{0.67\columnwidth}{!}{%
 \includegraphics{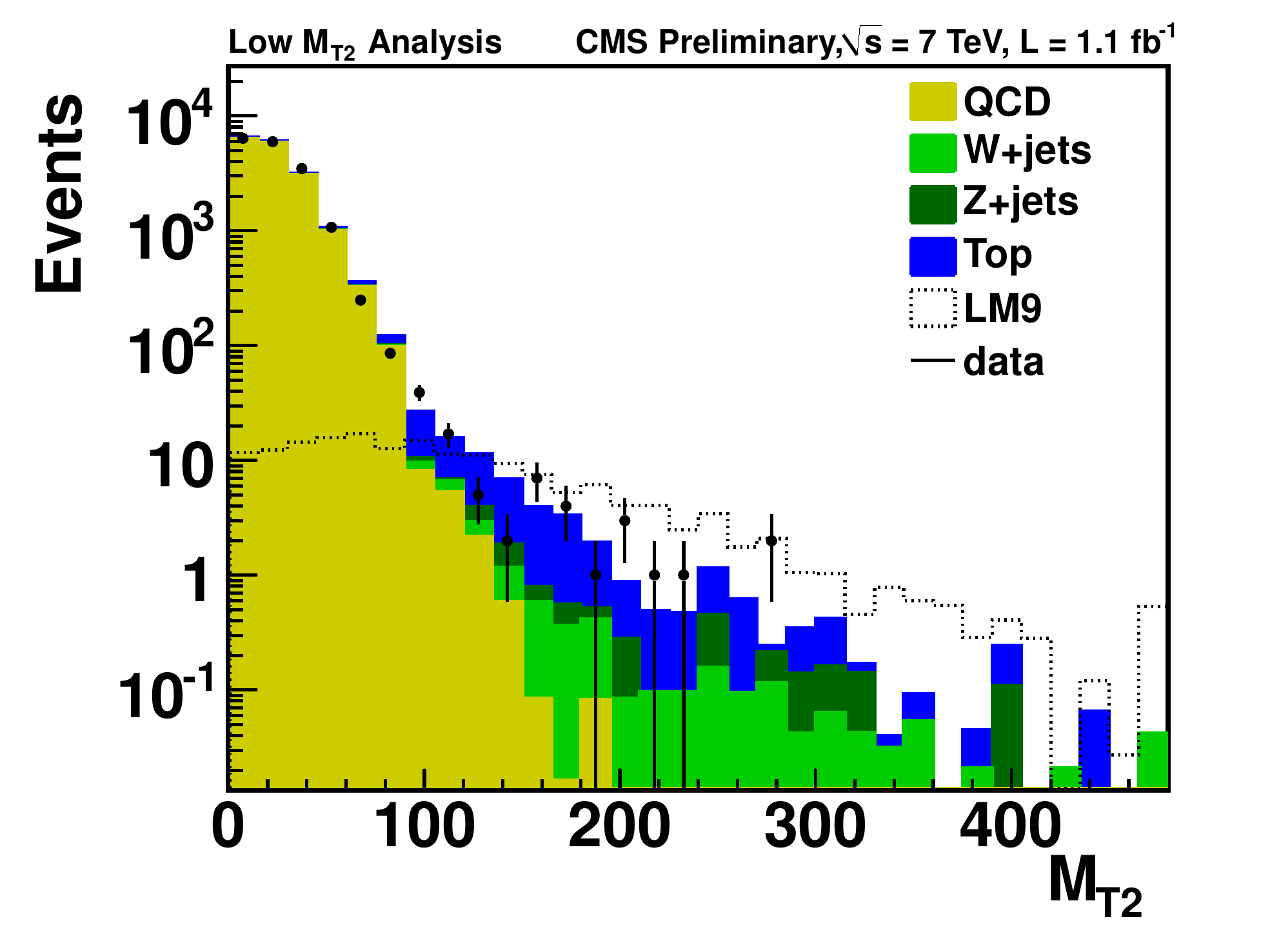} }
\caption{$M_{T2}$ distribution for the Low $M_{T2}$ selection. The MC background is normalized to 1.1 $\mathrm{fb}^{-1}$. A possible SUSY signal (LM9) is overlayed. Data are shown as dots on top of the background. }
\label{fig.lowMT2}       % Give a unique label
\end{center}
\end{figure}

The $M_{T2}$ distributions for the High $M_{T2}$ analysis and Low $M_{T2}$ analysis are shown in Figs. \ref{fig.highMT2} and \ref{fig.lowMT2}, table \ref{tab.results.highlowMT2} summarizes the results of the two analysis strategies. 
As no excess over background has been found limits have been set. 
\begin{table}[h!]
\begin{center}
\caption{Expected yield in the signal region from Standard Model (SM) background in simulation and from data-driven background predictions, as well as data yield for both analysis strategies. }
\label{tab.results.highlowMT2}       % Give a unique label
\begin{tabular}{lccc}
%\hline\noalign{\smallskip}
\hline\hline
                       & SM-MC & Background prediction                         & Data \\
\hline
High $M_{T2}$ & 10.6    & 12.6 $\pm$ 1.3 (stat) $\pm$ 3.5 (syst) & 12 \\
Low $M_{T2}$   & 14.3    & 10.6 $\pm$ 1.9 (stat) $\pm$ 4.8 (syst) & 19 \\
\hline\hline
\end{tabular}
\end{center}
\end{table}
\subsection{Exclusion Limits}
\label{seq.exclusions}
\begin{figure}[h!]
\begin{center}
\resizebox{0.67\columnwidth}{!}{%
 \includegraphics{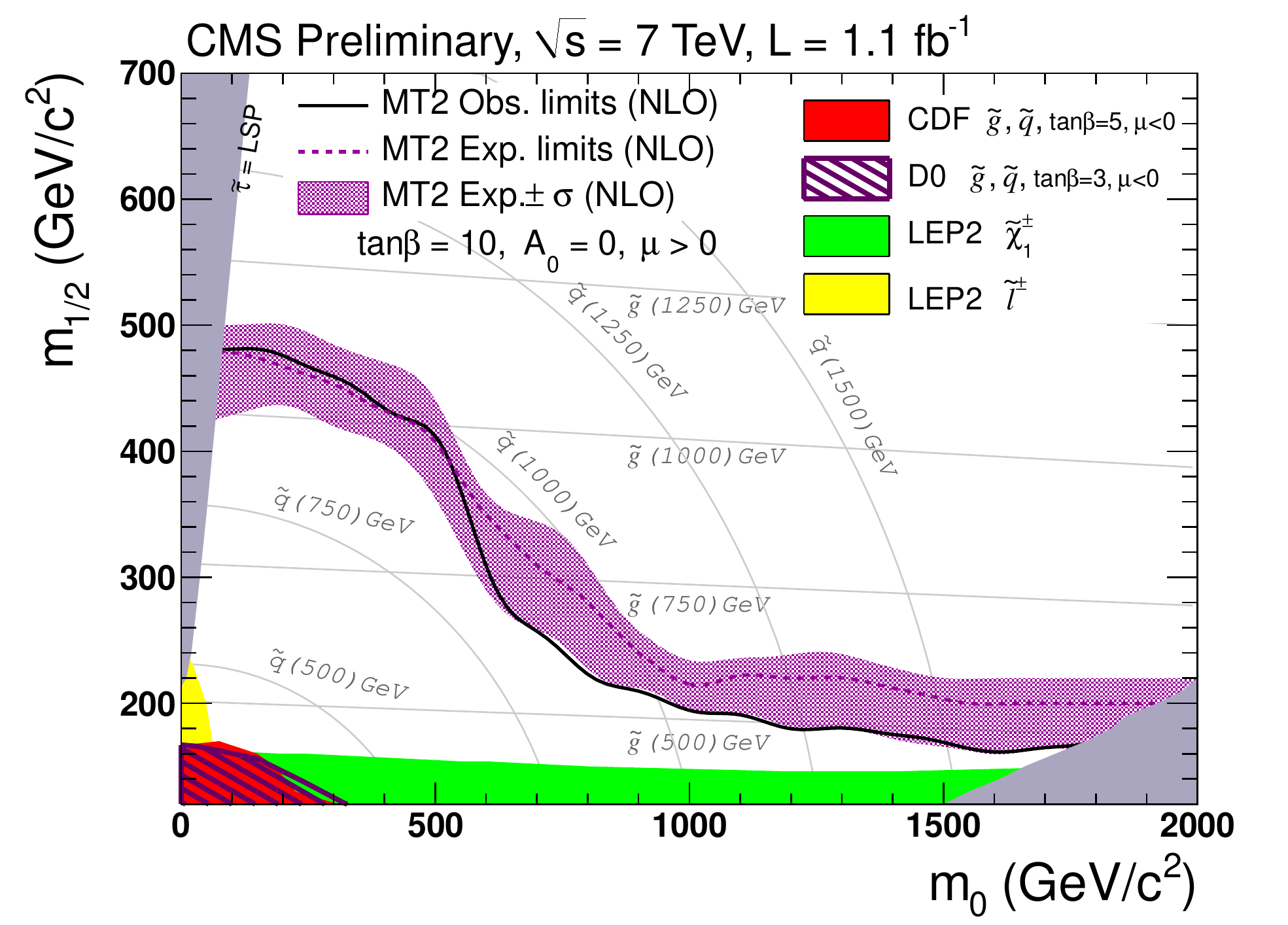} }
\caption{Combined exclusion limit in the mSUGRA/CMSSM ($\mathrm{m_0, ~m_{1/2}}$) plane with $\tan\beta$ = 10. }%In each point, the best of the two individual exclusions are taken. }
\label{fig.CMSSM.exclusion}       % Give a unique label
\end{center}
\end{figure}

\begin{figure}[h!]
\begin{center}
\resizebox{0.95\columnwidth}{!}{%
\includegraphics{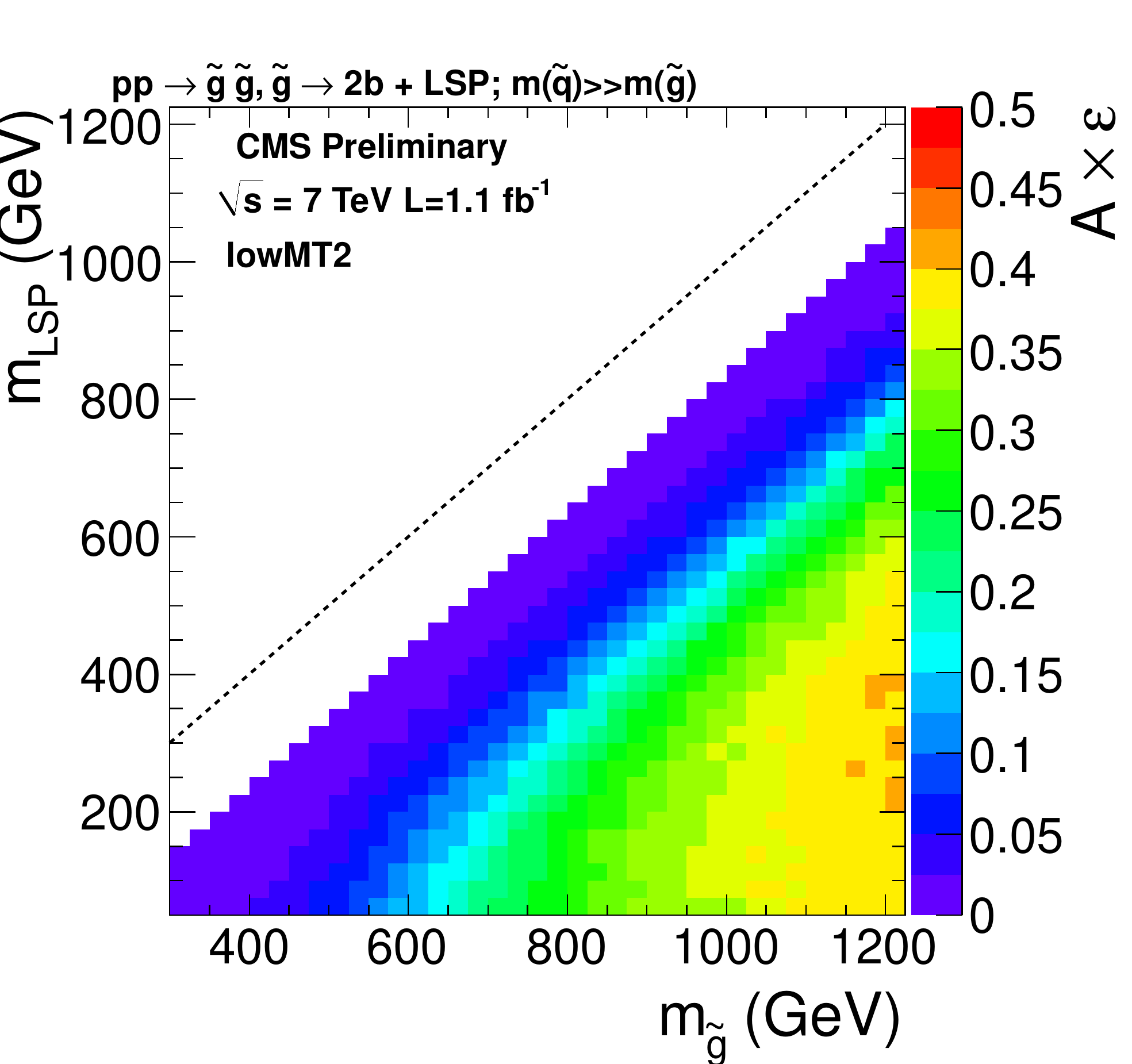}
 \includegraphics{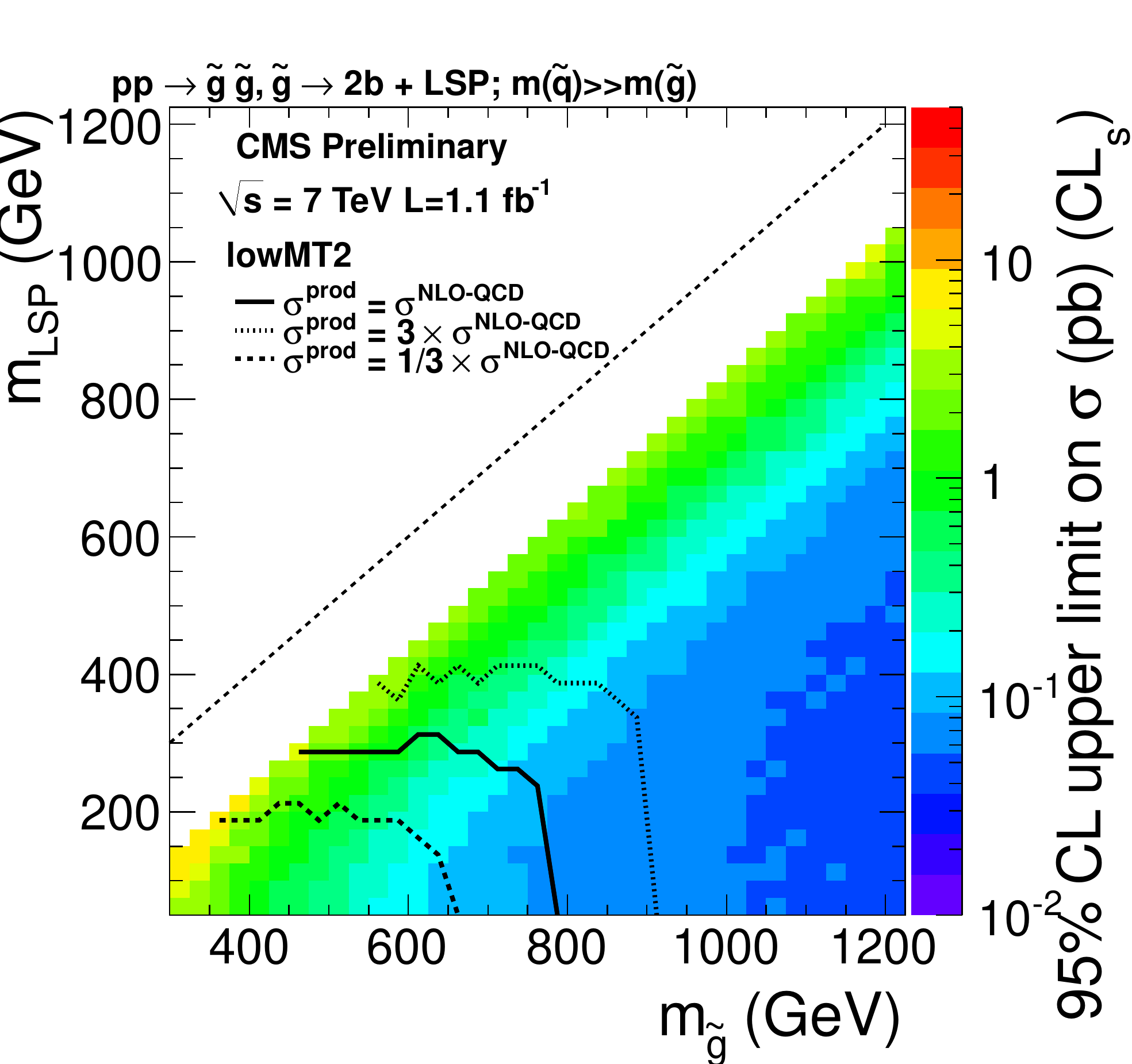} }
\caption{Model $pp\to\tilde{g}\tilde{g}, ~ \tilde{g}\to bb\tilde{\mathrm{\chi}}^0$ for the Low $M_{T2 }$ selection as a function of the mass parameters $\mathrm{m_{LSP}}$ and $\mathrm{m_{\tilde{g}}}$: Signal efficiencies (left), 95\% CL upper limit on cross section of the model (right).}
\label{fig.sms}
\end{center}
\end{figure}

First, model independent limits on $\sigma\times\mathrm{BR}$ within our acceptance has been derived by computing a 95\% upper limit on the number of events using a $\mathrm{CL_s}$ formulation \cite{PDG2010}. These limits are shown in table \ref{tab.limits.sigma.times.BR}.
\begin{table}[h!]
\begin{center}
\caption{Observed and expected limits on $\sigma\times\mathrm{BR}$ within the acceptance ot the two analysis strategies. }
\label{tab.limits.sigma.times.BR}       % Give a unique label
\begin{tabular}{lcc}
%\hline\noalign{\smallskip}
\hline\hline
                         &  \multicolumn{2}{c}{$\sigma \times$ BR (pb)} \\ 
                        &  observed limit    & expected limit   \\ 
\hline
High $M_{T2}$  &   0.010                &     0.011          \\
Low $M_{T2}$   &  0.020                 &      0.014           \\
\hline\hline
\end{tabular}
\end{center}
\end{table}
Exclusion limits at 95\% C.L. have been determined in the mSUGRA/CMSSM ($\mathrm{m_0}$, $\mathrm{m_{1/2}}$) plane. 
The results are shown in fig. \ref{fig.CMSSM.exclusion} for $\mathrm{A_0 =0}$, $\mathrm{\mu>0}$ and $\mathrm{\tan\beta = 10}$ combining the High and Low MT2 selections by taking the best expected limit in each point. 
Besides the exclusion in the mSUGRA/CMSSM plane the results are interpreted in a so-called Simplified Models topology. This is a simple signal model with exactly one decay mode which is only constrained by the kinematics and the masses of the participating particles. 
The Low $M_{T2}$ analysis is interpreted in a model where gluinos are pair produced and each gluino decays into two $b$-quarks and a neutralino, the LSP. 
The limits on the model cross sections and the signal efficiencies are shown in fig. \ref{fig.sms}.
\section{Conclusion}
\label{sec.conclusion}
We conducted a search for supersymmetry in hadronic final states using the $M_{T2}$ variable calculated from massless pseudojets. 
A data set containing 1.1 $\mathrm{fb}^{-1}$ of integrated luminosity in $\sqrt{s}$ = 7 TeV $pp$ collisions recorded by the CMS detector during the 2011 LHC run was analyzed. 
Two complementary analyses were performed to probe a larger SUSY phase space. 
In both analyses the tail of the $M_{T2}$ is sensitive to a possible SUSY signal. 
As no evidence for a signal was found, we set upper limits on the cross section times branching ratio within our acceptance. 
Exclusions limits were established in the mSUGRA/CMSSM parameter space, as well as in a Simplified Model topology.

\end{document}